\documentclass[%
 reprint,
 amsmath,amssymb,
 aps,
prapplied,
]{revtex4-2}

\usepackage{graphicx}
\usepackage{dcolumn}
\usepackage{bm}
\usepackage{mathrsfs}

\begin{document}

\preprint{}

\title{Non-local superconducting single-photon detector}

\author{Federico Paolucci}
\email{federico.paolucci@pi.infn.it}
\affiliation{INFN Sezione di Pisa, Largo Bruno Pontecorvo, 3, I-56127 Pisa, Italy}

\date{\today}

\begin{abstract}
We present and theoretically analyse the performance of an innovative \emph{non-local} superconducting single-photon detector. The device operates thanks to the \emph{energy-to-phase} conversion mechanism, where the energy of the absorbed single-photon is transformed in a variation of the superconducting phase.
To this scope, the detector is designed in the form of a double-loop superconductor/normal metal/superconductors ($SNS$) Josephson interferometer, where the detection occurs in a \emph{long} $SNS$ junction and the read-out is operated by a \emph{short} $SNS$ junction.
The variation of the superconducting phase across the read-out junction is measured by recording the quasiparticle current flowing through a tunnel coupled superconducting probe.
By exploiting realistic geometry and materials, the detector is able to reveal single-photons of frequency down to 10 GHz when operated at 10 mK. Furthermore, the device provides value of signal-to-noise ratio up $10^{4}$ in the range 10 GHz-10 THz by selecting the magnetic flux and the bias voltage. 
This device can find direct applications as single-photon detector in both basic science and quantum technology, while the energy-to-phase conversion mechanism can be at the basis of non-local read-out and memory architectures for superconducting qubits.

\end{abstract}

\maketitle


\section{\label{sec:Intro}Introduction}

Superconducting local detectors are the state-of-the-art for sensing electromagnetic radiation of energies ranging from hundreds of keV (X-rays) to tens of $\mu$eV ($\sim10$ GHz). 
In particular, superconductivity is exploited to boost the sensitivity of both bolometers and calorimeters operating at the lower frequency boundary. 
Indeed, the most used superconducting detectors in the GHz-THz band are the transition-edge sensors ($TES$s) \cite{Irwin1995} and the kinetic inductance detectors ($KID$s) \cite{Day2003}, due to their extreme sensitivity and robustness. 
$TES$ bolometers reached a noise equivalent power ($NEP$) on the order of $10^{-19}$ W/Hz$^{1/2}$ \cite{Suzuki2014}, while $KID$s showed a $NEP$ of $\sim10^{-18}$ W/Hz$^{1/2}$ \cite{deVisser2014}.  
Moreover, these structures showed a frequency resolution of several hundreds of GHz when operated as single-photon detectors \cite{Niwa2017,Elefante2023}. To push detection technology towards higher sensitivities and lower frequencies, miniaturization of the sensing element was ridden \cite{Paolucci2020}.  
Indeed, superconducting nanowire single-photon detectors ($SNSPD$s) have been developed in the visible and infrared bands by exploiting quasi one-dimensional channels \cite{Natarajan2012}. 
Furthermore, Josephson effect in hybrid superconducting nano-structures is a fundamental knob to design innovative radiation sensors and improve the detection sensitivity. 
Indeed, single-photon counters based on tunnel Josephson junctions working down to a few tens of GHz \cite{Oelsner2013,Guarcello2019}, ultrasensitive proximity nanobolometers \cite{Giazotto2008,Kokkoniemi2019}, thermoelectric calorimeters \cite{Tero2018,Paolucci2023}, and the current-tunable Josephson escape sensor ($JES$) \cite{Paolucci2020b} have been designed and realized. 
On the one hand, the strong temperature dependence of superconductivity increases the sensitivity of a detector. On the other hand, a local read-out operation performed directly on the sensing element could alter the detection, due to the introduction of even a small overheating. To avoid this drawback, a \emph{non-local} architecture separating the detector and the read-out element can be beneficial. To this scope, detectors exploiting the temperature-to-phase ($TPC$) \cite{Virtanen2018} and temperature-to-voltage ($TVC$) \cite{Solinas2018} conversion have been proposed.

Here, we propose a non-local superconducting single-photon detector based on the energy-to-phase conversion mechanism. To this end, we exploit a double-loop Josephson interferometer \cite{Clarke2004,Paolucci2022}, where the absorption of a single-photon in a \emph{long} superconductor/normal metal/superconductor ($SNS$) Josephson junction (detector) causes the variation of the superconducting phase across a \emph{short} $SNS$ junction (read-out). 
The output signal is obtained by measuring the quasiparticle current flowing through a tunnel probe coupled directly to the read-out junction. The double-loop geometry and the asymmetry between the two junctions (detector and read-out) ensure optimal energy-to-phase conversion, thus boosting the detection sensitivity. Indeed, our detector is expected to efficiently reveal single-photons of frequency down to 10 GHz. Furthermore, the detector can be in-situ tuned by controlling both the magnetic flux piercing the interferometer and the bias voltage. 

The paper is organized as follows: Section \ref{sec:Device} introduces the device structure and its operation principle; Section \ref{sec:Materials} lists the geometrical dimensions and the materials constituting each element of the structure;  Section \ref{sec:Temperature} shows the effects of the photon-absorption on the electronic temperature of the detector junction; Section \ref{sec:Phase} evaluates the energy-to-phase conversion and its effects on the energy spectrum of the $SNS$ read-out junction; Section \ref{sec:Curr} shows the dependence of the output current on the photon frequency; Section \ref{sec:Perf} displays the detection performance of the device; and Section \ref{sec:Concl} resumes the concluding remarks. 
 
\section{\label{sec:Device}Operation principle and structure}

\begin{figure} [t!]
    \begin{center}
    \includegraphics [width=\columnwidth]{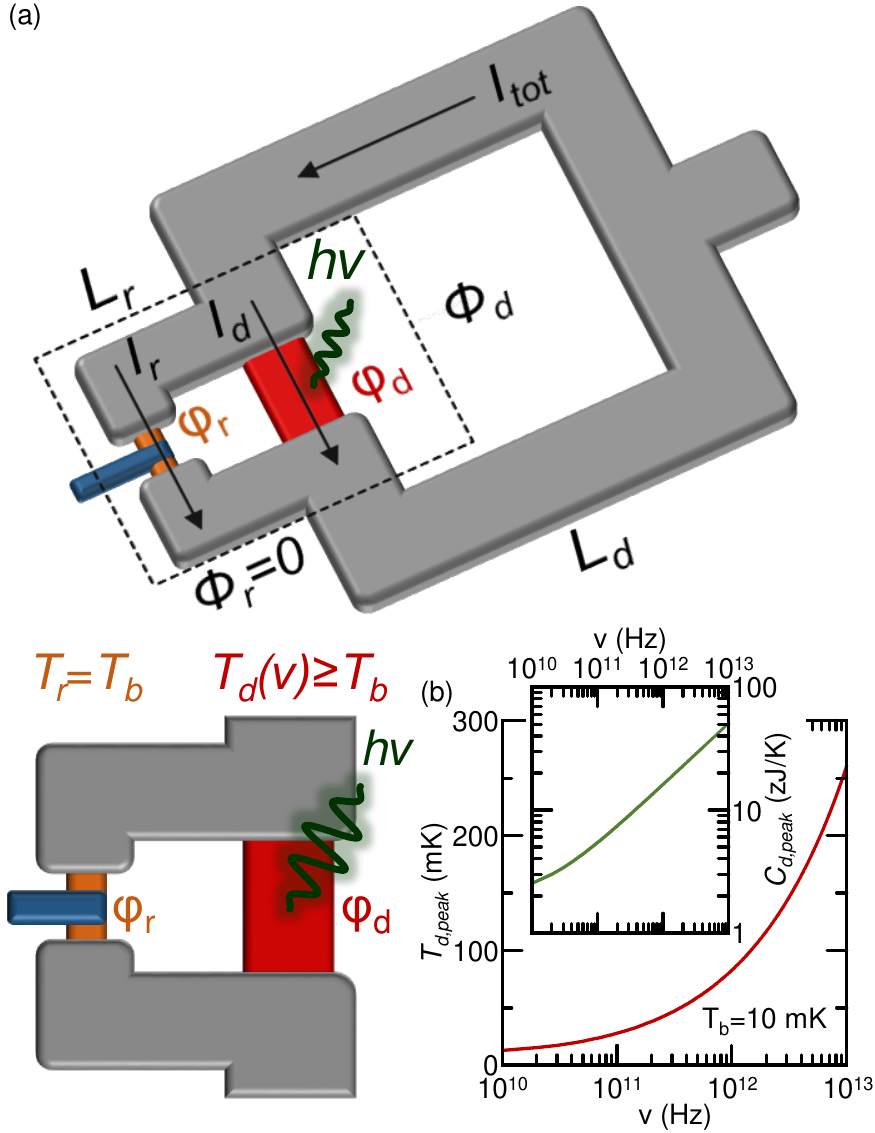}
    \end{center}
    \caption{\textbf{(a)} Schematic of the non-local superconducting single-photon detector. The device is structured as a long detection ($d$, red) and a short read-out ($r$, orange) SNS Josephson junction interrupting two superconducting loops (grey) of inductance $L_d$ and $L_r$, respectively. The total circulating current ($I_{tot}$) divides between $d$ (with switching current $I_d$) and $r$ (with switching current $I_r$). The first loop is pierced by a magnetic flux $\Phi_d$, while the read-out loop is fully screened form the external magnetic field ($\Phi_r=0$). Thus, the two junctions show the same drop of the superconducting phase ($\varphi_d=\varphi_r$). A photon of frequency ($\nu$) is absorbed by $d$ thus increasing its temperature $T_d$, while the temperature of $r$ stays constant at the bath temperature ($T_r=T_b$). The related change of $\varphi_r$ is recorded by a superconducting probe (blue) tunnel-coupled to $r$.
    \textbf{(b)} Peak temperature of the detection junction ($T_{d,peak}$) versus frequency ($\nu$). Inset: Thermal capacitance of $d$ versus $\nu$ calculated at the peak temperature. All calculations were performed for a bath temperature $T_b=10$ mK.}
\label{Fig1}
\end{figure}

The structure of our single-photon detector is shown in Fig. \ref{Fig1}a, where a \emph{long} detection $SNS$ junction ($d$, red) is integrated in a two-loops Josephson interferometer (grey) together with a \emph{short} $SNS$ read-out Josephson junction ($r$, orange). Then, a superconducting probe ($p$, blue) is tunnel coupled directly to $r$.
Its operation is based on the strong temperature dependence of the supercurrent flowing in $d$, since it is in the \emph{long}-junction limit \cite{Golubov2004}. Indeed, the current-to-phase relation (CPR) of $d$ can be written as
\begin{equation}
    I_{d} (T_d,\varphi_d)=I_{C,d}(T_d) \sin{\left(\varphi_d\right)},
    \label{Icrit}
\end{equation}
where $I_{C,d}$ is the critical current of the junction, $T_d$ is the electron temperature of $d$, and $\varphi_d$ the superconducting phase drop across the junction. The absorption of a single-photon causes the increase of $T_d$ (see Section \ref{sec:Temperature} for details) and, thus, the exponential damping of the critical current of $d$ \cite{Zaikin1981,Wilhelm1997}
\begin{equation}
\begin{aligned}    
 &I_{C,d}(T_d)=\frac{64 \pi k_B T_d}{(3+2\sqrt{2})eR_d}\times \\
 &\sqrt{\frac{2\pi k_B T_d}{E_{th,d}}}\exp{\bigg(-\sqrt{{\frac{2\pi k_B T_d}{E_{th,d}}}}\bigg)},
 \end{aligned}
    \label{AmbeEq}
\end{equation}
where $e$ is the electron charge, $k_B$ is the Boltzmann constant, $R_d$ is the normal-state resistance of the junction, $E_{th,d}=\hbar D_d/\mathcal{L}_d^2$ is the Thouless energy (with $\hbar$ the reduced Planck constant, $D_d$ the diffusion coefficient and $\mathcal{L}_d$ the length of the proximitized normal metal strip). Since $d$ is inserted in a superconducting loop, it is subject to fluxoid quantization
\begin{equation}
     \varphi_d+\frac{2\pi L_d}{\Phi_0}I_{tot}=\frac{2\pi\Phi_d}{\Phi_0},
     \label{Ffconv}
\end{equation}
where $L_d$ is the inductance of the detection loop, $I_{tot}=I_d+I_r$ (with $I_r$ the switching current of the read-out junction) is the total supercurrent circulating in the detector, $\Phi_d$ is the magnetic flux bias, and $\Phi_0$ is the flux quantum. As a consequence, the increase of $T_d$ due to the absorption of a single-photon of energy $h\nu$ (with $h$ the Planck constant and $\nu$ the photon frequency) causes a variation of $\varphi_d$. Therefore, our single-photon detector operates thanks to \emph{energy-to-phase} conversion. 

The \emph{non-local} read-out is guaranteed by the second superconducting loop, which is assumed to be fully screened from the external magnetic flux ($\Phi_r=0$). Consequently, the phase drop across $r$ ($\varphi_r$) is linked to $\varphi_d$ by
\begin{equation}
          \varphi_r-\varphi_d+\frac{2\pi L_r}{\Phi_0}I_{r}=0,
          \label{eq:phase-locking}
\end{equation}
where $L_r$ is the inductance of the read-out loop. Thus, the absorption of a single-photon by $d$ indirectly produces the variation of $\varphi_r$ (see Section \ref{sec:Phase} for details). The typical method to record a phase variation in a $SNS$ junction is the tunnel spectroscopy of the normal metal island, thus to implement a superconducting quantum interference proximity transistor (SQUIPT) \cite{Giazotto2010}. The choice of $r$ in the short-junction limit ensures the maximum response to changes of $\varphi_r$ of $I_r$, and thus of its density of states (DoS) \cite{Giazotto2011}. Within this assumption, the CPR of $r$ takes the Kulik-Omel'yanchuk (KO) form \cite{Kulik1975}
\begin{equation}
\begin{aligned}
    &I_{r} (T_r,\varphi_r)=\frac{\pi\Delta(T_r)}{eR_r}\cos{\left( \frac{\varphi_r}{2}\right)}\;\times\\
    &\int\limits_{\Delta(T_r)\cos{\left( \frac{\varphi_r}{2}\right)}}^{\Delta(T_r)}\mathrm{d}\varepsilon\;\frac{\tanh{\big(\frac{\varepsilon}{k_BT_r}\big)}}{\sqrt{\varepsilon^2-\Delta^2(T_r)\cos^2{\left(\frac{\varphi_r}{2}\right)}}},
    \end{aligned}
     \label{KO}
\end{equation}
where $\Delta(T_r)$ is the temperature-dependent energy gap of the ring, $R_r$ is the normal-state resistance of $r$, and $\varepsilon$ is the energy relative to the chemical potential of the superconductor. Accordingly, $\varphi_r$ strongly influences the DoS of the normal metal stripe ($\mathcal{N}_{r}$) forming the read-out junction \cite{Artemenko1979,Heikkila2002}
\begin{equation}
        \begin{aligned}
        &\mathcal{N}_{r}(x,\varepsilon,\varphi_r)=\text{Re}\Bigg[\sqrt{\frac{(\varepsilon+i\Gamma_r)^2}{(\varepsilon+i\Gamma_r)^2-\Delta^2(T_r)\cos^2(\frac{\varphi_r}{2})}}\;\times\\
        &\cosh\left( \frac{2x-\mathcal{L}_r}{\mathcal{L}_r}\text{arccosh}\sqrt{\frac{(\varepsilon+i\Gamma_r)^2-\Delta^2(T_r)\cos^2(\frac{\varphi_r}{2})}{(\varepsilon+i\Gamma_r)^2-\Delta^2(T_r)}}\right)\Bigg],
        \end{aligned}
        \label{eq:DoS}
\end{equation}
where $\Gamma_r$ is the Dynes broadening parameter \cite{Dynes1984} and $x\in [0,\mathcal{L}_r]$ is the spatial coordinate along $r$. Furthermore, the superconducting minigap induced by the proximity effect  is written \cite{DeGennes}
\begin{equation}
    E_{g,r}(T_r,\varphi_r)=\Delta(T_r)\cos{\left(\frac{\varphi_r}{2}\right)}.
    \label{eq:minigap}
\end{equation}
Consequently, both $\mathcal{N}_{r}(T_r,\varphi_r)$ and $E_{g,r}(T_r,\varphi_r)$ strongly depend on $\varphi_r$ (at fixed $T_r$) and thus on the electronic temperature of the detection junction.

To maximise the responsivity of the detector, we implement the output tunnel probe $p$ in the form of a superconducting wire. Indeed, SQUIPTs with superconducting tunnel probes show enhanced signals \cite{Jabdaraghi2014,Virtanen2016}. Thus, the normalized DoS of $p$ reads
\begin{equation}
\mathcal{N}_{p}(\varepsilon,V)=\left|\text{Re}\left[\frac{(\varepsilon-eV+i\Gamma_p)}{\sqrt{(\varepsilon-eV+i\Gamma_p)^2-\Delta_p^2(T_p)}}\right]\right|,
\end{equation}
where $\Gamma_p$ is the Dynes parameter, $\Delta_p$ is the temperature-dependent superconducting energy gap, $T_p$ the electronic temperature and $V$ the voltage drop across the tunnel barrier. The resulting quasiparticle tunnel current reads
\cite{Giazotto2011}
\begin{equation}
\begin{aligned}
           &I(\varphi_r,V)=\frac{1}{ew_pR_p}\int\limits_{\frac{\mathcal{L}_r-w_p}{2}}^{\frac{\mathcal{L}_r+w_p}{2}}\text{d}x\;\times\\
           &\int\limits_{-\infty}^{\infty}\text{d}\epsilon\;\mathcal{N}_{r}(x,\epsilon,\varphi_r)\mathcal{N}_{p}(\epsilon, V)\big[f_{0,p}(\varepsilon-eV)-f_{0,r}(\varepsilon)\big],
           \end{aligned}
           \label{Eq:currtunn}
\end{equation}
where $w_p$ is the width of $p$, $R_p$ is the normal-state resistance the $r-p$ tunnel junction, while $f_{0,r}$ and $f_{0,p}$ are the Fermi distributions of $r$ and $p$, respectively. Equation \ref{Eq:currtunn} clearly shows that the output current is strongly dependent on $\varphi_r$ and, thus, on the electron temperature of the detector $T_d$.

\section{\label{sec:Materials}Materials}
In this section, we introduce the physical dimensions and the materials chosen for each element of the proposed non-local single-photon detector. In particular, we exploit a structure feasible with standard nano-fabrication techniques. These structural features will be employed to determine the detection performance of the device at a bath temperature $T_b=10$ mK.

The superconducting ring is assumed to be made of aluminum ($\Delta_0=200\;\mu$eV), The detection and read-out ring are assumed to show an inductance $L_d=100$ pH and $L_r=10$ pH,  and respectively. 

The copper $SNS$ detector Josephson junction (DoS at the Fermi energy $N_{F,d}=1.56\times10^{47}$ J$^{-1}$m$^{-3}$ and diffusion constant $D_d=10^{-2}$ m$^2$s$^{-1}$) is characterized by a length $\mathcal{L}_d=1.5\;\mu$m and a volume $\mathscr{V}_d=12\times10^{-21}$ m$^{-3}$. Consequently, the normal-state resistance of $d$ is $R_d=\rho_d\mathcal{L}_d^2/\mathscr{V}_d=\mathcal{L}_d^2/(e^2N_{F,d}D_d\mathscr{V}_d)=28\;\Omega$.
Furthermore, the Thouless energy takes the value $E_{Th,d}=2.93\;\mu\text{eV}\ll\Delta_0$, thus the junction lies in the long-limit regime.

The read-out short $SNS$ Josephson junction is made of copper ($N_{F,r}=1.56\times10^{47}$ J$^{-1}$m$^{-3}$ and $D_r=10^{-2}$ m$^2$s$^{-1}$), too. It is assumed to show a normal-state resistance $R_r=500\;\Omega$ and a Dynes parameter $\Gamma_r=10^{-4}\Delta_0$.

Finally, the output probe ($p$) is made of alluminum ($\Delta_{0,p}=200\;\mu$eV and $\Gamma_p=10^{-4}\Delta_{0,p}$) and it is coupled to $r$ by a tunnel connection of normal-state resistance $R_p=10\;\text{k}\Omega$.

\section{\label{sec:Temperature}Energy-to-temperature conversion}
Here, we evaluate the increase of the temperature of the detector junction $T_d$ due to the absorption of a photon of energy $h\nu$. We assume that a suitable antenna funnels the energy of the single-photons directly to $d$, since their direct absorption by $d$ is limited by their long wavelength. The energy released by the single-photons can be confined in the normal region of the detection junction by implementing Andreev mirrors (AMs) \cite{Andreev1964}, which are superconducting elements with energy gap larger than the ring ($\Delta_{0,AM}\gg\Delta_0$) preventing heat out-diffusion.
Under this condition, the absorption of a photon increases exclusively the temperature of $d$ above the phonon temperature ($T_b$, i.e. the bath temperature). In particular, the temperature rises instantaneously up to a maximum value $T_{d,peak}$ and then the electron-phonon collisions tend to bring back the equilibrium ($T_d=T_b$) \cite{Giazotto2006}.
Thus, we derive the dependence of $T_{d,peak}$ and the electron heat capacitance ($C_{d}$) on the frequency of the incident photon ($\nu$). To this end, we introduce the temperature-dependent thermal capacitance of the normal-state defined as \cite{Solinas2018}
\begin{equation}
    C_{d}(T_d)=\frac{\pi^2}{3}\mathscr{V}_d N_{F,d} k_B^2 T_d,
    \label{Eq:capac}
\end{equation}
where $N_{F,d}$ is the density of states at the Fermi energy of $d$. 
We overestimate the thermal capacitance (and thus underestimate the temperature increase), since the presence of a superconducting minigap induced by proximity effect damps $C_d$ \cite{Rabani2008,Rabani2009}. Moreover, the thermal capacitance in a long Josephson junction changes over the length of the normal metal wire (since the minigap is position dependent) \cite{leSeur2008}, thus complicating the analysis. We stress that our choice provides a simple solution without influencing the operation principle of the detector and providing an underestimated sensitivity.
Within this model, the temperature $T_{d,peak}$ can be calculated by solving \cite{Moseley1984,Chui1992}
\begin{equation}
    h\nu= \int_{T_{b}}^{T_{d,peak}} \text{d}T_d\; C_{d}=\frac{\pi^2}{6} \mathscr{V}_d N_{F,d} k_B^2\big( T_{d,peak}^2 - T_{b}^2 \big),
\end{equation}
since the electron-electron thermalization is much faster than all the other energy exchange mechanisms in the system. Indeed, the peak temperature takes the form
\begin{equation}
    T_{d,peak}(\nu)=\sqrt{T_b^2+\frac{6h\nu}{\pi^2\mathscr{V}_dN_{F,d}k_B^2}},
    \label{Eq:temp}
\end{equation}
that is the peak temperature increases above $T_b$ with the square root of the photon frequency. This behavior is shown in Fig. \ref{Fig1}b, where $T_{d,peak}$ is calculated as a function of $\nu$ at $T_b=10$ mK. In particular, a single-photon of frequency 10 GHz increases the electronic temperature of $d$ to about 13 mK, while a 10 THz photon generates a $T_{d,peak}\simeq260$ mK.

Finally, by substituting Eq. \ref{Eq:temp} in Eq. \ref{Eq:capac}, we can compute the dependence of the thermal capacitance of $d$ on the incident photon frequency
\begin{equation}
\begin{aligned}
    &C_{d,peak}(\nu)=\frac{\pi^2}{3} \mathscr{V}_d N_{F,d} k_B^2 \sqrt{T_b^2+\frac{6h\nu}{\pi^2\mathscr{V}_dk_B^2}}=\\
    &=\pi k_B \sqrt{\frac{\mathscr{V}_d N_{F,d}}{3}\bigg(  \frac{\pi^2 \mathscr{V}_d N_{F,d} k_B^2}{3} T_b^2+2h\nu \bigg)}.
    \end{aligned}
\end{equation}
We stress that this is the value of heat capacitance of $d$ at the photon absorption, thus it depends on the square root of $\nu$, as shown in the inset of Fig. \ref{Fig1}b. Indeed, by increasing $\nu$ of 3 orders of magnitude, the thermal capacitance rises of about 20 times.  

\section{\label{sec:Phase}Energy-to-phase conversion}
\begin{figure} [t!]
    \begin{center}
    \includegraphics [width=\columnwidth]{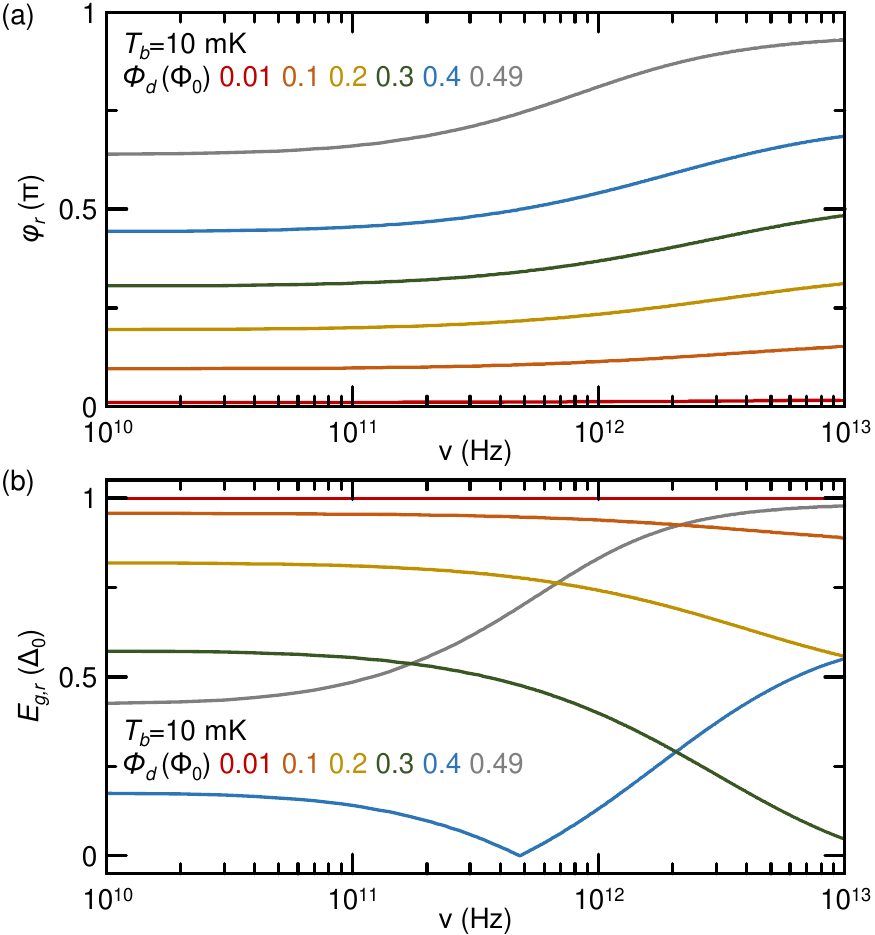}
    \end{center}
    \caption{\textbf{(a)} Superconducting phase drop across the read-out junction ($\varphi_r$) versus the frequency of the incident photon ($\nu$) calculated for different values of the magnetic flux biasing the device ($\Phi_d$). 
    \textbf{(b)} Superconducting minigap induced in $r$ ($E_{g,r}$) versus $\nu$ calculated for different values of $\Phi_d$. All calculations were performed at $T_b=10$ mK.}
\label{Fig2}
\end{figure}

Here, we evaluate the variations of $\varphi_r$ due to absorption of a photon of frequency $\nu$. To this end, we combine the temperature and phase dependence of the electronic transport of Josephson interferometers (see Sec. \ref{sec:Device}) with the energy-to-temperature conversion mechanism in $d$ (see Sec. \ref{sec:Temperature}).

At a given value of $T_b$ and $\Phi_d$, the absorption of a photon generates a change of the phase bias of $r$ given by 
\begin{equation}
     \begin{aligned}
        &\varphi_r(\nu)=\\
        &=\frac{2\pi}{\Phi_0}\bigg[\Phi_d -L_dI_{C,d}(\nu)\sin \bigg(\varphi_r+2\pi\frac{L_rI_r(T_b,\varphi_r)}{\Phi_0} \bigg)- \\
        &(L_d+L_r)I_r(T_b,\varphi_r) \bigg],
    \end{aligned}
    \label{Eq:PhiRvsNUmAIN}
\end{equation}
where the photon frequency dependent critical current of $d$ takes the form
\begin{equation}
\begin{aligned}    
 &I_{C,d}(\nu)=\frac{\sqrt{4096\pi^2 k_B^2T_b^2+\frac{24576h\nu}{\mathscr{V}_dN_{F,d}}}}{(3+2\sqrt{2})eR_d}\;\times \\
 &\sqrt[4]{\frac{4\pi^2 k_B^2T_b^2+\frac{24h\nu}{\mathscr{V}_dN_{F,d}}}{E_{th,d}^2}}\;\times\\
 &\exp{\Bigg(-\sqrt[4]{\frac{4\pi^2 k_B^2T_b^2+\frac{24h\nu}{\mathscr{V}_dN_{F,d}}}{E_{th,d}^2}}\Bigg)}.
 \end{aligned}
    \label{ICdNumAIN}
\end{equation}
As a consequence, $\varphi_r$ is predicted to show a strong dependence on the incident photon frequency. The details of the calculations for deriving Eqs. \ref{Eq:PhiRvsNUmAIN} and \ref{ICdNumAIN} are provided in Appendix \ref{App:Phase}.

Figure \ref{Fig2}a shows the dependence of the phase drop across the read-out junction on the photon frequency for different values of $\Phi_d$. In particular, the variation of $\varphi_r$ with $\nu$ increases with the flux bias approaching its maximum for $\Phi_d\to 0.5\Phi_0$. Indeed, the detector shows $\delta\varphi_r(\Phi_d=0.01\Phi_0)=|\varphi_r(0.01\Phi_0 , 10\; \text{GHz})-\varphi_r(0.01\Phi_0,10\; \text{THz})|\simeq5.7\times10^{-3}\pi$ and $\delta\varphi_r(\Phi_d=0.49\Phi_0)\simeq0.29\pi$. Interestingly, the phase starts to vary at lower frequencies by increasing $\Phi_d$, thus suggesting a larger sensitivity of the single-photon detector. In the same flux range, $\varphi_d$ reaches values larger than $\pi/2$. This implies that the value of $\Phi_d$ determines whether the photon absorption causes an increase of decrease of the induced minigap of $r$ (see Eq. \ref{eq:minigap}). 
Figure \ref{Fig2}b shows the dependence of the induced minigap $E_{g,r}$ on $\nu$ for different values of flux bias. The minigap monotonically increases for $\Phi_d\leq0.3\Phi_0$, while $E_{g,r}$ is damped for $\Phi_d=0.49\Phi_0$ (grey curve). Differently, the minigap shows a non-monotonic dependence on $\nu$ for $\Phi_d=0.4\Phi_0$. In fact, in this case $\varphi_r$ crosses $\pi/2$ (see Fig. \ref{Fig2}a), which defines the periodicity of $E_{g,r}(\varphi_r)$.
Consequently, the choice of a specific value of $\Phi_d$ defines the sensitivity of the system to the absorption of single-photon of different energy.

\section{\label{sec:Curr}Output current versus $\nu$}
\begin{figure} [t!]
    \begin{center}
     \includegraphics [width=\columnwidth]{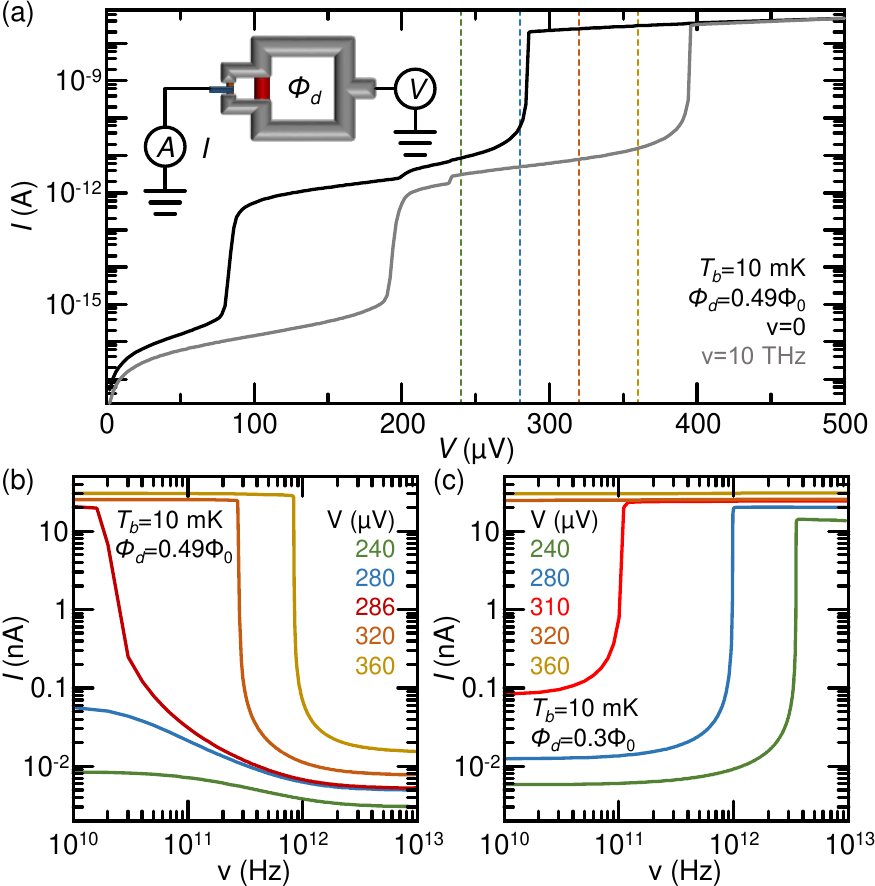}
    \end{center}
    \caption{\textbf{(a)} Current ($I$) versus voltage ($V$) characteristics of the non-local single-photon detector in the absence of incident radiation (grey) and for an incident photon of frequency $\nu=10$ THz calculated at $\Phi_d=0.49\Phi_0$. The dashed vertical lines indicate suitable values of the bias voltage for the device operation. Inset: scheme of the read-out scheme of the detector, where $\Phi_d$ is the bias flux, $V$ is the bias voltage and $I$ is the measured quasiparticle tunnel current.
    \textbf{(b)} $I$ versus $\nu$ calculated for different values of $V$ at $\Phi_d=0.49\Phi_0$. $V=286\;\mu$V (red) is the bias that ensures current modulation at lower frequency.
    \textbf{(c)} $I$ versus $\nu$ calculated for different values of $V$ at $\Phi_d=0.3\Phi_0$. $V=310\;\mu$V (red) is the bias that ensures current modulation at lower frequency. All calculations were performed at $T_b=10$ mK.}
\label{Fig3}
\end{figure}

In this section, we evaluate influence of the single-photon absorption on the quasiparticle transport of the non-local detector. Indeed, the energy-to-phase conversion implies a frequency-dependent change of the density of states of $r$ (see Eq. \ref{eq:DoS}) and thus a modification of the detector output signal (see Eq. \ref{Eq:currtunn}).

Figure \ref{Fig3}a shows the quasiparticle current ($I$) versus voltage ($V$) characteristics of the detector calculated by solving Eq. \ref{Eq:currtunn} at $T_b=10$ mK and $\Phi_d=0.49\Phi_0$. In particular, we evaluated the difference between the response in the absence of radiation ($\nu=0$, black) and in the presence of a single-photon of frequency $\nu=10$ THz (grey) by exploiting Eqs. \ref{eq:DoS}, \ref{Eq:PhiRvsNUmAIN} and \ref{ICdNumAIN} to evaluate $\varphi_r$. The $IV$ curves show the typical behavior of a tunnel junctions between two superconductors, where the current dramatically increases at the transition to the normal-state ($V\sim[\Delta_p(T_b)+E_{g,r}(\varphi_r)]/e$). As shown in Fig. \ref{Fig2}b, at $\Phi_d=0.49\Phi_d$ the absorption of a photon causes the increase of the induced minigap in $r$. Thus, the transition to the normal-state behavior of the $IV$ characteristics moves towards larger values of the bias voltage after photon absorption. This feature is combined with a change of the sub-gap quasiparticles conduction. Therefore, we can measure the absorption of a single photon by recording the tunnel current at a fixed $V$ (see Inset of Fig. \ref{Fig3}a). In general, appropriate bias voltages lie in the range $\Delta_0\leq eV \leq2\Delta_0$, since $\Delta_p(10\;\text{mK})\sim\Delta_0$ and $0\leq E_{g,r} (\varphi_0) \leq \Delta_0$. In our setting, the voltage bias range is thus $200\;\mu\text{V}\leq V \leq 400\;\mu\text{V}$. This allows to generate easily measurable modulations of $I$. Despite $I$ is also modulated for $eV<\Delta_0$ ($V<200\;\mu\text{V}$), the absolute value of the current is rather small and thus unpractical in concrete experimental setups. Indeed, we chose four values of $V$ in the best bias range to determine the current output and the performance of the non-local single-photon detector (see dashed lines in Fig. \ref{Fig3}a).

We first focus on the modulation of $I$ for values of the magnetic flux bias guaranteeing a monotonic dependence of $E_{g,r}$ on $\nu$ (increasing or decreasing). Accordingly, the quasiparticle current tends to decrease after the photon absorption for $\Phi_d=0.49\Phi_0$ (Figure \ref{Fig3}b), while $I$ rises for $\Phi_d=0.3\Phi_0$ (Figure \ref{Fig3}c). Beyond the values of voltage shown in \ref{Fig3}a, we included the $I(\nu)$ characteristics calculated for values of $V$ ensuring the onset of the current variation at the lowest photon frequency (red lines). In particular, the highest current modulation is ensured for $V=286\;\mu$V at $\Phi_d=0.49\Phi_0$ and for $V=310\;\mu$V at $\Phi_d=0.3\Phi_0$. 
At $\Phi_d=0.49\Phi_0$, the current starts to vary of several orders of magnitude starting from $\nu\simeq20$ GHz, while for a flux bias $\Phi_d=0.3\Phi_0$ the strong $I$ increase happens around 70 GHz. This behavior is direct consequence of the $E_{g,r}(\nu)$ characteristics shown in Fig. \ref{Fig2}b and suggest sensitivity of the detector also at those low frequencies.

\begin{figure} [t!]
    \begin{center}
            \includegraphics [width=\columnwidth]{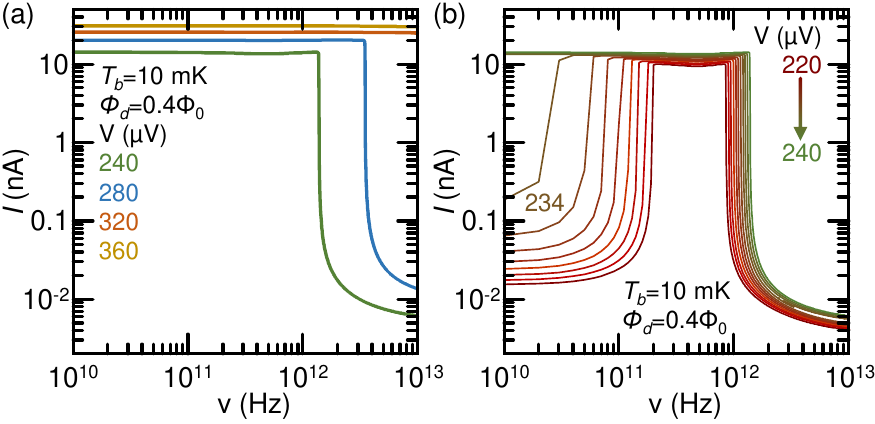}
    \end{center}
    \caption{\textbf{(a)} $I$ versus $\nu$ calculated for different values of bias voltage ($V$).
    \textbf{(b)} $I$ versus $\nu$ calculated for values $V$ highlighting the non-monotonic modulation of $E_{g,r}$ with increasing incident photon frequency. All calculations were performed at $\Phi_d=0.4\Phi_0$ and $T_b=10$ mK.}
\label{Fig4}
\end{figure}

We now switch our attention on $\Phi_d=0.4\Phi_0$, where the $E_{g,r}(\nu)$ characteristics is non-monotonic. Thus, we expect a richer and more complex behavior of the detector output current. Indeed, the sign of the derivative of $I$ changes with increasing $\nu$ (see Eq. \ref{Eq:currtunn}). For $V\geq240\;\mu$V, the current monotonically decreases with the radiation frequency (see Fig. \ref{Fig4}a), since the $E_{g,r}(\nu)$ characteristics is single-valued and monotonically increasing with $\nu$ at the energies corresponding to these values of $V$, as shown by the blue curve in Fig. \ref{Fig2}b. Differently, for lower values of bias voltage, the $E_{g,r} (\nu)$ is non-monotonic thus providing two possible values of $\nu$ at a given $V$. Consequently, the $I(\nu)$ characteristics are non-monotonic for $V<240\;\mu$V. In particular, the current increases at low frequencies, stay constant for about a decade in frequency, and starts to decrease again at larger values of $\nu$ (see Fig. \ref{Fig4}b). On the one hand, the onset of the $I$ dampening is weakly dependent on $V$ (from $\sim1$ THz at $V=220\;\mu$V to $\sim2$ THz at $V=240\;\mu$V). On the other hand, the beginning of the current increases is tuned over more than one order of magnitude by changing the bias voltage in the same range. In particular, by setting $V=234\;\mu$V the current is already sensitive for $10$ GHz single-photons, while $V=220\;\mu$V provides a strong current modulation at about 200 GHz. 

To conclude, the choice of the flux $\Phi_d$ and bias voltage $V$ allow to decide the sign of the current variation with photon absorption and to choose the sensitivity frequency range.

\section{\label{sec:Perf}Single-photon detection performance}
Here, we investigate the sensitivity of our non-local single-photon detector by calculating the typical figures of merit for a calorimeter, such as the signal-to-noise ratio ($SNR$), the energy resolution ($h\nu/\delta E$) and the time constant ($\tau_{1/2}$).

\begin{figure} [t!]
    \begin{center}
    \includegraphics [width=\columnwidth]{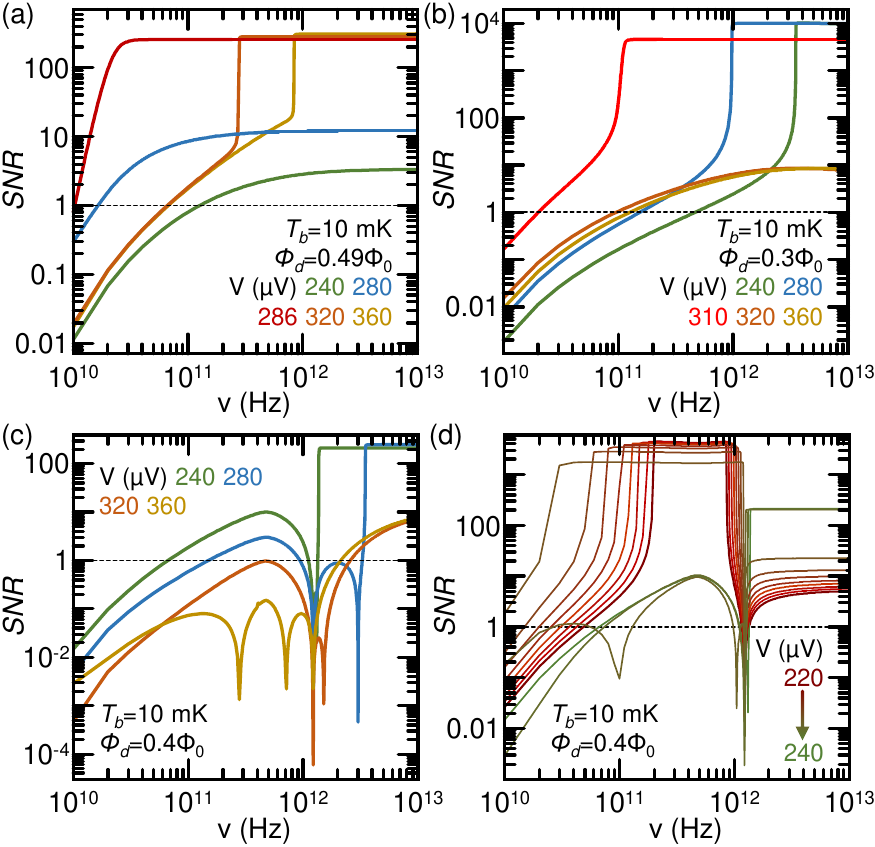}
    \end{center}
    \caption{\textbf{(a)} Signal-to-noise ratio ($SNR$) versus photon frequency ($\nu$) calculated for different values of bias voltage ($V$) at $\Phi_d=0.49\Phi_0$.
    \textbf{(b)} $SNR$ versus $\nu$ calculated for different values of $V$ at $\Phi_d=0.3\Phi_0$.
    \textbf{(c)} $SNR$ versus $\nu$ calculated for different values of voltage ($V$) at $\Phi_d=0.4\Phi_0$.
    \textbf{(d)} $SNR$ versus $\nu$ calculated for values of $V$ around the maximum sensitivity at $\Phi_d=0.4\Phi_0$. 
    The dashed lines indicate the lower limit for single-photon detection ($SNR=1$). All curves are calculated for $T_b=10$ mK.}
\label{Fig5}
\end{figure}

The signal-to-noise ratio evaluates the magnitude of the signal produced by the detector when sensing a single-photon with respect to the background noise. In voltage bias, the $SNR$ reads
 \begin{equation}
     SNR\,(\nu)=\frac{|I[V,\varphi_r(\nu)]-I[V,\varphi_r(\nu=0)]|}{\sqrt{S_{I,tot}}\sqrt{\omega}},
 \end{equation}
where $S_{I,tot}$ is the total current-noise spectral density (see Appendix \ref{App:Noise} for details), and $\omega =1$ MHz is the measurement bandwidth. This choice satisfies the requirement $\omega \geq 2 \pi/\tau_{1/2}$, where $\tau_{1/2}$ is the characteristics time constant of the detector (as calculated below). 

Figure \ref{Fig5} shows the dependence of the $SNR$ on the photon frequency calculated at the same values of $\Phi_d$ and $V$ selected for the $I(\nu)$ characteristics (see Sec. \ref{sec:Curr}). 
In particular, panel a shows the $SNR\,(\nu)$ characteristics calculated for $\Phi_d=0.49\Phi_0$. As expected for this flux bias, the $SNR$ monotonically increases with the photon frequency. In addition, the detector is sensitive to low frequency single-photons in the conditions of flux and voltage bias showing the larger change of probe current. Indeed, the detector is able detect single-photons of frequency $\sim10$ GHz ($SNR>1$) for $V=286\;\mu$V with maximum values of signal-to-noise ratio of about 255 at $\nu=10$ THz.
Similarly, Fig. \ref{Fig5}b shows the $SNR\,(\nu)$ characteristics obtained for $\Phi_d=0.3\Phi_0$. On the one hand, the detector is less sensitive to low frequency photons ($SNR\geq1$ for $\nu\gtrsim20$ GHz at $V=310\;\mu$V). On the other hand, the detector reaches $SNR\sim 10^4$ at 10 THz for $V=240\;\mu$V, since the output current increases when absorbing a single-photon, thus providing a signal much larger than the dark current noise ($\nu=0$).

Differently, the $SNR$ is non-monotonic with the frequency at $\Phi=0.4\Phi_0$ (see Fig. \ref{Fig5}c-d), thus reflecting the behavior of the $I(\nu)$ characteristics. Interestingly, the detector is insensitive to low frequency photons for a wide range of bias voltages ($SNR>1$ for $\nu\gtrsim2.5$ THz for $V=360\;\mu$V). In the optimally biased case ($V=234\;\mu$V, Fig. \ref{Fig5}d), the detector is able to reveal 10 GHz single-photons, but it becomes blind for frequencies $\nu\sim1$ THz. Therefore, the bias voltage can be exploited to tune the sensitivity of the detector to single-photons of the desired energy while \emph{filtering} part of the remaining electromagnetic spectrum.
We note that the sensitivity of the detector to photons of extremely low frequency (down to 10 GHz) balloons the importance of the proposed non-local read-out mechanism. Indeed, low dissipated power into the detector junction  due to direct tunnel measurements or even the quasiparticles created by dispersive measurements can affect the detector response. As a consequence, these effects can decrease the effective sensitivity of the detector by increasing the noise levels and the dark-counts.  

\begin{figure} [t!]
    \begin{center}
    \includegraphics [width=\columnwidth]{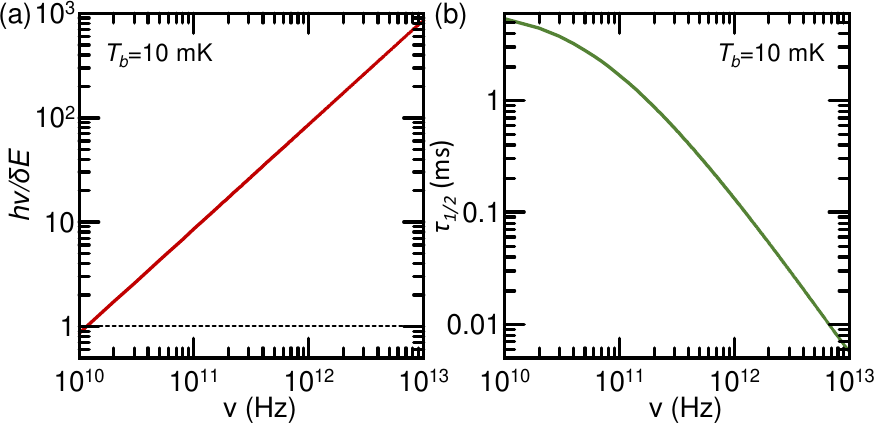}
    \end{center}
    \caption{\textbf{(a)} Energy resolution ($h\nu/\delta E$) versus photon frequency ($\nu$). The dashed line indicates $h\nu/\delta E=1$.
    \textbf{(b)} Time constant ($\tau_{1/2}$) versus $\nu$.
    All curves are calculated at $T_b=10$ mK.}
\label{Fig7}
\end{figure}

The energy resolution describes the ability of a detector to distinguish the energy of the revealed single-photon. For the proposed device, it can be evaluated through
\begin{equation}
    \frac{h\nu}{\delta E}(\nu,T_b)=\frac{h\nu}{4\sqrt{2\ln{2}\,k_BT_b^2C_d(T_d=T_b)}},
\end{equation}
where the thermal capacitance $C_d$ is evaluated at $T_b$, that is at the temperature of $d$ before the photon absorption. Figure \ref{Fig7}a shows ${h\nu}/\delta E$ for our non-local detector calculated at $T_b=10$ mK. In particular, the sensor can discern single-photons of energy $\sim10$ GHz (${h\nu}/\delta E \geq 1$). Moreover, the detector shows an energy resolution of about $10^3$ for 10 THz single-photons. We stress that we exploited the normal-state thermal capacitance (see Eq. \ref{Eq:capac}), thus underestimating the energy resolution. This underestimation is larger for low energies, where the overheating is lower and the difference between the value of $C_d$ in the superconducting and normal-state is maximum \cite{Rabani2008,Rabani2009}. Therefore, these are low boundary values of the energy resolution. 

Finally, the characteristic time constant of a single-photon detector can be determined with the electron-phonon relaxation halftime, that is the time necessary to have $T_d=(T_{b}+T_{d,peak})/2$. For our device, it is defined
\begin{equation}
    \tau_{1/2}=\int_{(T_{b}+T_{d,peak})/2}^{T_{d,peak}} \text{d}T_d \; \frac{C_{d}(T_d)}{P_{e,ph}(T_d,T_b)}, 
\end{equation}
where $P_{e,ph}=\mathscr{V}_d\Sigma_d(T_d^5-T_b^5)$ (with $\Sigma_d=2\times10^9$ Wm$^{-3}$K$^{-5}$ the e-ph coupling constant of copper) is the electron-phonon thermalization in $d$ \cite{Giazotto2006}. We note that, to be consistent with the heat capacitance, we consider the e-ph thermalisation in the normal-state. In our device, the characteristic time of the detector decreases by increasing $\nu$ (see Fig. \ref{Fig7}b) from $\sim6$ ms at 10 GHz to $\sim70\;\mu$s at 10 THz. This is due to the rise fo the e-ph thermalisation at the large electronic temperatures, as reached at high single-photon frequency. In addition, we point out that the thermalisation is dominated by the electron-phonon scattering even in the absence of AMs, since an aluminum thin film is almost completely thermally insulating up to elecronic temperatures of about 280 mK \cite{Paolucci2020} (which is larger than $T_{d,peak}\simeq270$ mK reached for 10 THz single-photons).

\section{\label{sec:Concl}Conclusions}
In summary, we propose a non-local superconducting single-photon detector that operates thank to the \emph{energy-to-phase} conversion mechanism. The device is composed of a double-loop interferometer with a long $SNS$ Josephson junction acting as detector and a short $SNS$ operating as read-out. The device output signal is the quasiparticle current recorded by a superconducting electrode tunnel-coupled directly to the read-out junction. Differently from previous non-local detectors, our analysis takes into account the non-ideal phase bias due to the finite kinetic inductance of the superconducting loops composing the device. By exploiting experimentally feasible structure and materials, the proposed detector is able to reveal single-photons of frequency down to 10 GHz with an energy resolution larger than 1 when operated at 10 mK. Furthermore, the sensitivity can be tuned by controlling the magnetic flux piercing the detection loop and the bias voltage. In particular, the response to single-photons in specific windows of the electromagnetic range could be filtered by setting the biasing parameters.
Since the large detector sensitivity arises from the exponential temperature dependence of the critical current of the long $SNS$ detector Josephson junction, the use of large critical temperature materials (such as Nb or NbN) could allow to increase the operation temperature. In these systems, the detection sensitivity could be slightly limited by the increase of the electron-phonon thermalization at higher phonon temperatures. 

This single-photon detector can find applications in both quantum technologies and basics science, such as quantum cryptography \cite{Gisin2002,Tittel2019}, optical quantum computing \cite{Obrien2007}, THz spectroscopy \cite{Peiponen2013} and particle search \cite{Sikivie1983}. Furthermore, the non-local and non-invasive detection scheme might also be exploited for the readout \cite{Siddiqi2021} or memory \cite{Ligato2021} operations of superconducting qubits.

\begin{acknowledgments}
The author thanks G. Lamanna, S. Roddaro, G. Signorelli, P. Spagnolo, A. Tartari and A. Tredicucci for fruitful discussions. The author acknowledge F. Bianco for critically reading the manuscript.

The author declares no competing financial interest.
\end{acknowledgments}

\appendix

\section{\label{App:Phase}Calculation of $\varphi_r$ versus $\nu$}

Here, we derive the dependence of the superconducting phase drop across the read-out junction ($\varphi_r$) on the frequency of the incident photon ($\nu$). To this scope, we need to consider the fluxoid quantization of the detection loop \cite{Clarke2004}
\begin{equation}
     \varphi_d+\frac{2\pi L_d}{\Phi_0}I_{tot}=2\pi \frac{\Phi_d}{\Phi_0},
     \label{FLuxQuantd}
\end{equation}
the fluxoid quantization of the read-out loop (assuming $\Phi_r=0$)
\begin{equation}
     \varphi_r-\varphi_d+\frac{2\pi L_r}{\Phi_0}I_{r}=0,
     \label{FLuxQuantr}
\end{equation}
and the conservation of the circulating current
\begin{equation}
    I_{tot}=I_d+I_r.
    \label{ConsCurr}
\end{equation}
We can extract $\varphi_r$ by combining Eq. \ref{FLuxQuantd} and Eq. \ref{FLuxQuantr} 
\begin{equation}
     \begin{aligned}
        &\varphi_r=\varphi_d-2\pi\frac{L_rI_{r}}{\Phi_0}=\\
        &=2\pi\frac{\Phi_d}{\Phi_0}-2\pi\frac{L_dI_{tot}}{\Phi_0}-2\pi\frac{L_rI_{r}}{\Phi_0}=\\
        &=\frac{2\pi}{\Phi_0}\big[\Phi_d- L_dI_{tot}-L_rI_{r} \big].
    \end{aligned}
\end{equation}
By substituting the circulating current conservation (Eq. \ref{ConsCurr}), the phase drop can be written as
\begin{equation}
     \begin{aligned}
        &\varphi_r=\frac{2\pi}{\Phi_0}\big[\Phi_d- L_d(I_d+I_r)-L_rI_{r} \big]=\\
        &=\frac{2\pi}{\Phi_0}\big[\Phi_d- L_dI_d-L_dI_r-L_rI_{r} \big]=\\
        &=\frac{2\pi}{\Phi_0}\big[\Phi_d- L_dI_d-I_r(L_d+L_r) \big].
    \end{aligned}
\end{equation}
We now simplify $\varphi_d$ and introduce the dependence of $I_{d}$ on temperature. To this end, we introduce the expression of the switching current of $d$ (long $SNS$ Josephson junction)
\begin{equation}
    I_d(T_d,\varphi_d)=I_{C,d}(T_d)\sin(\varphi_d)
\end{equation}
and substitute
\begin{equation}
    \varphi_d=\varphi_r + 2\pi \frac{L_rI_r(T_r,\varphi_r)}{\Phi_0}.
\end{equation}
As a consequence, at a given value of $\Phi_d$, the phase drop across $r$ strongly depends on both $T_d$ and $T_r$, as shown by
\begin{equation}
     \begin{aligned}
        &\varphi_r(T_d,T_r)=\\
        &=\frac{2\pi}{\Phi_0}\bigg[\Phi_d -L_dI_{C,d}(T_d)\sin(\varphi_d)-(L_d+L_r)I_r(T_r,\varphi_r) \bigg]=\\
        &=\frac{2\pi}{\Phi_0}\bigg[\Phi_d -L_dI_{C,d}(T_d)\sin \bigg(\varphi_r+2\pi\frac{L_rI_r(T_r,\varphi_r)}{\Phi_0} \bigg)- \\
        &(L_d+L_r)I_r(T_r,\varphi_r) \bigg].
    \end{aligned}
\end{equation}
We note that $\varphi_r$ has an exponential dependence on $\sqrt{T_d}$, since $I_{C,d}\propto \exp(-\sqrt{T_d})$ for a Josephson junction in the long-diffusive limit (see Eq. \ref{AmbeEq} of the main text) \cite{Zaikin1981,Wilhelm1997}.
The absorption of a photon increases the temperature of $d$ above $T_b$ (see Eq. \ref{Eq:temp} and Fig. \ref{Fig1}b) up to the maximum value $T_{d,peak}$ (see Sec. \ref{sec:Temperature} of the main text), while it does not affect the read-out junction temperature ($T_r=T_b$). Therefore, at a given $T_b$ and $\Phi_d$, $\varphi_r$ is only a function of $\nu$, as shown by
\begin{equation}
     \begin{aligned}
        &\varphi_r(\nu,T_b)=\\
        &=\frac{2\pi}{\Phi_0}\bigg[\Phi_d -L_dI_{C,d}(\nu)\sin \bigg(\varphi_r+2\pi\frac{L_rI_r(T_b,\varphi_r)}{\Phi_0} \bigg)- \\
        &(L_d+L_r)I_r(T_b,\varphi_r) \bigg].
    \end{aligned}
    \label{Eq:PhiRvsNu}
\end{equation}
Finally, by substituting Eq. \ref{Eq:temp} into Eq. \ref{AmbeEq} of the main text, we obtain the dependence of the critical current of $d$ on the photon frequency (Eq. \ref{ICdNumAIN} of the main text)
\begin{equation}
\begin{aligned}    
 &I_{C,d}(\nu)=\frac{\sqrt{4096\pi^2 k_B^2T_b^2+\frac{24576h\nu}{\mathscr{V}_dN_{F,d}}}}{(3+2\sqrt{2})eR_d}\times \\
 &\sqrt[4]{\frac{4\pi^2 k_B^2T_b^2+\frac{24h\nu}{\mathscr{V}_dN_{F,d}}}{E_{th,d}^2}}\times\\
 &\exp{\Bigg(-\sqrt[4]{\frac{4\pi^2 k_B^2T_b^2+\frac{24h\nu}{\mathscr{V}_dN_{F,d}}}{E_{th,d}^2}}\Bigg)}.
 \end{aligned}
    \label{ICdNuApp}
\end{equation}

\newpage
\section{\label{App:Noise}Current-noise spectral density}
The total current-noise spectral density takes the form
\begin{equation}
    S_{I,tot}=S_{I,TFN}+S_{I,p},
\end{equation}
where $S_{I,TFN}$ is the current noise relate to the thermal fluctuations in the detect and $S_{I,p}$ is the low-frequency current-noise of $p$ calculated in the idle state (that is in the absence of radiation).

The thermal fluctuation noise can be calculated by \cite{Virtanen2018}
\begin{equation}
  S_{I,TFN}\simeq\frac{|I[V=0,\varphi_r(\nu)]-I[V=0,\varphi_r(\nu=0)]|^2}{\omega},
\end{equation}
which is related to the different between the current fluctuations in the presence and in the absence of radiation when no voltage bias is applied.

The current noise of the probe takes the form

\begin{equation}
    S_{I,p}=2eI[V,\varphi_r(\nu=0)]\coth{\bigg( \frac{eV}{2k_BT_b} \bigg)},
\end{equation}
which is the intrinsic noise of the tunnel probe at the voltage $V$ and temperature $T_b$, that is in a dark environment.

\end{document}